\documentclass[a4paper,11pt]{article}
\usepackage{jcappub} 

\usepackage{graphicx}
\usepackage{caption}
\usepackage{subcaption}
\usepackage{bm}
\usepackage{comment}

\usepackage{siunitx}
\let\svqty\qty
\usepackage{physics}
\let\qty\svqty
\usepackage{amsmath,amssymb,amsfonts}
\def\be#1\ee{\begin{align}#1\end{align}} 
\def\bse#1\ese{\begin{subequations}#1\end{subequations}}
\newcommand{\qu}[1]{``#1''}
\usepackage[scr=boondoxo,scrscaled=1.05]{mathalfa}

\usepackage[capitalize]{cleveref}
\crefname{equation}{Eq.}{Eqs.}
\crefname{figure}{figure}{figures}
\crefname{table}{table}{tables}
\crefname{subequation}{Eqs.}{Eqs.}
\labelcrefformat{subequation}{#2(#1)#3}
\crefformat{subequations}{#2(#1)#3}
\crefname{section}{Sec.}{Sec.s}
\crefname{appendix}{App.}{Apps.}

\usepackage[dvipsnames]{xcolor}

\definecolor{rRGB}{RGB}{169.2, 155.0, 0.5}
\definecolor{rRGB2}{RGB}{139.2, 95.0, 50}
\usepackage{soul}

\author{Raúl Carballo-Rubio$^{1}$ and}
\affiliation{$^{1}$Instituto de Astrof\'isica de Andaluc\'ia (IAA-CSIC),
Glorieta de la Astronom\'ia, 18008 Granada, Spain}
\emailAdd{raul.carballorubio@iaa.csic.es}
\author{Jacopo Mazza$^{2}$}
\affiliation{$^{2}$Université Paris-Saclay, CNRS/IN2P3, IJCLab, 91405 Orsay, France}
\emailAdd{jacopo.mazza@ijclab.in2p3.fr}

\title{\boldmath Exploring the construction of field equations for regular black holes with dynamical regularization scales}

\abstract{The study of the dynamics of regular black holes is an emergent area of research. Recent works have discussed how to deform the Einstein field equations to regularize spherically symmetric black hole solutions due to the introduction of a new physics scale acting as regulator. In this exploratory paper, we consider for the first time the problem of constructing field equations in which the regularization scale is a dynamical field. Such field equations may provide an effective description of the evolution of the cores of regular black holes due to the backreaction of matter, and a novel avenue for constructing alternative dynamical solutions of theoretical and phenomenological interest. We discuss a general family of field equations describing spherically symmetric gravity coupled to a scalar field, and study whether the sector with a constant scalar field contains solutions describing static regular black holes with no additional physics scales aside from the ones provided by the scalar field.
We show that this question can be reduced to solving a linear partial differential equation for the coupling functions between the scalar and the metric degrees of freedom.}

\begin{document}
\maketitle
\flushbottom

\section{Introduction}\label{sec:Intro}

One of the most striking consequences of the Einstein field equations is the necessary existence of singularities inside of black holes for physically reasonable matter contents~\cite{Penrose:1964wq,Senovilla:2014gza}, and of Cauchy horizons whenever charge or angular momentum are present~\cite{Hawking:1973uf}. These problems are a strong driving force for the development of non-singular completions of general relativity, spanning diverse theoretical and phenomenological developments (see the recent review~\cite{Carballo-Rubio:2025fnc} and references therein).

While constructing a complete picture of black hole interiors in known theoretical frameworks is currently out of reach, it may be possible to extract qualitative insights from the study of effective models. In particular, here we will be interested in effective models in which spacetime singularities are excised, resulting in regular spacetimes. The study of the so-called \emph{regular black holes} has a long history that started with works by Bardeen, Frolov, Dymnikova and Hayward, among others~\cite{Bardeen1968,Frolov:1981mz,Dymnikova:1992ux,Hayward:2005gi}, and still provide the best-motivated alternative to classical black holes. As such, regular black holes are the subject of a vibrant area of research, from the classification of singularity-free spacetimes~\cite{Carballo-Rubio:2019nel,Carballo-Rubio:2019fnb,Carballo-Rubio:2021wjq,Mazza:2021rgq,Franzin:2022wai,Carballo-Rubio:2022nuj,LimaJunior:2025uyj,Eichhorn:2025pgy} to the study of their phenomenology~\cite{Carballo-Rubio:2018jzw,Cardoso:2019rvt,Held:2019xde,Eichhorn:2021etc,Franzin:2022iai,Guerrero:2022qkh,Eichhorn:2022oma,Eichhorn:2022bbn,Carballo-Rubio:2022aed,Carballo-Rubio2023,Olmo:2023lil,Cardoso:2023guh,Mazza:2023mvx,Koch:2025gaw}.

Regular black holes require the introduction of at least one new physical scale $\ell$ controlling the magnitude of deviations with respect to general relativity as well as the structure of the resulting regular spacetime. For simplicity, $\ell$ is generally considered to be a new fundamental scale, typically taking Planckian values (although in some phenomenological studies the latter condition is relaxed, e.g.,~\cite{Held:2019xde}). The size of the internal regular core in these deformations shrinks to zero in the limit $\ell\rightarrow0$, in which the solutions and predictions of general relativity are recovered.

In this paper, we will go beyond these approaches and consider theoretical frameworks in which the regularization scale can be dynamical. This is motivated by the observation that the dynamical behavior of regular black holes is largely unknown, while there are arguments and calculations indicating that matter perturbations, both classical and semiclassical, lead to large dynamical effects around inner horizons~\cite{Brown:2011tv,Frolov:2016gwl,Frolov:2017rjz,Carballo-Rubio:2018pmi,Bonanno:2020fgp,Carballo-Rubio:2021bpr,Carballo-Rubio:2024dca,Boyanov:2025otp}, which has been proposed to lead to the \qu{inflation} of the inner horizon~\cite{Barcelo:2020mjw,Barcelo:2022gii} and be connected to black-hole-to-white-hole transitions~\cite{Barcelo:2014npa,Barcelo:2014cla} (a connection that has been recently gaining attention~\cite{Barenboim:2024dko,Arrechea:2026rua}). While a full dynamical picture of such processes has been evasive, it is not unreasonable to think that the resulting evolution can be effectively described by a time-dependent (and possibly position-dependent) length scale $\ell$ encapsulating the leading dynamical behavior of the regular core in the presence of perturbations.

Our discussion will build upon recent breakthroughs in the development of phenomenological field equations containing regular black holes as vacuum solutions. This approach started, to the best of our knowledge, with the discussion of effective equations for spherically symmetric gravitational collapse based on 2-dimensional dilaton gravity performed by Ziprick and Kunstatter~\cite{Ziprick:2010vb} (see also~\cite{Louis-Martinez:1993agn,Taves:2014laa}), which was further extended for vacuum solutions in~\cite{Kunstatter:2015vxa}; these results were summarized and extended in \cite{ColleauxNonpolynomialLagrangian2018,Colleaux:2019ckh,ColleauxRationalRegular2026}. In the vacuum case, the field equations can be recast in terms of 2-dimensional Horndeski theory, which has been leveraged to define a set of 4-dimensional master field equations generalizing the spherically symmetric Einstein field equations~\cite{Carballo-Rubio:2025ntd} and to describe exact solutions of 4-dimensional field equations describing dynamical regular black holes~\cite{Boyanov:2025pes}, as well as several other applications~\cite{BorissovaEffectiveGeometrodynamics2026,BorissovaRegularBlack2026a,BorissovaAll2D2026,BorissovaModifiedFriedmann2026,Arrechea:2026ngi,Borissova$g_ttg_rr1$2026,Carballo-Rubio:2026mvj,MazzaBlackBounces2026,BorissovaFormationExtremal2026,Thaalba:2026abz}. The construction of such effective theories has also been independently discussed from a Hamiltonian perspective~\cite{Alonso-BardajiSpacetimeGeometry2024,Alonso-BardajiDynamicalTheory2025,Zhang:2025ccx}. This effective approach to the study of spherically symmetric spacetimes can be straightforwardly generalized to arbitrary dimensions, and has an interesting overlap with the study of regular black hole solutions in quasitopological theories defined in 5 or more dimensions~\cite{Myers:2010ru,Oliva:2010eb,Bueno:2016xff,Hennigar:2017ego,Ahmed:2017jod,Bueno:2019ycr} which, due to their reduction to specific 2-dimensional Horndeski theories for spherically symmetric spacetimes, implies that any spherically symmetric solution such as the ones discussed in~\cite{BuenoRegularBlack2025c,BuenoRegularBlack2025b,Bueno:2024eig,Bueno:2024zsx,Bueno:2025gjg} is equivalent to those obtained in the effective approach. This is a remarkable convergence of independent developments which, while focusing on the same problem, originated from quite different starting points. 

Here, we will further the development of this framework by considering a biscalar extension of 2-dimensional Horndeski gravity. Specifically, we will investigate the possibility of promoting the regularization scale $\ell$ to a function of the theory's dynamical degrees of freedom --- which constitutes, to the best of our knowledge, an unexplored problem. Being a first exploration of this subject, we will focus on defining a general family of field equations, identifying in a second stage specific subsets in which these equations simplify. In some restricted subsets, we will be able to determine that black hole solutions with dynamical regularization scales do not arise, while in more general cases we present general equations that must be satisfied for the problem to have a solution.

This paper is organized as follows. We start with a detailed discussion of the field equations we will be working with in Sec.~\ref{sec:fieldequations}. After making contact with previous works in the single-field case in Sec.~\ref{sec:singext}, we study situations in which the additional scalar field becomes constant in Sec.~\ref{sec:const_scalar}. We provide a concluding discussion in Sec.~\ref{sec:discussion}.

\section{Field equations}\label{sec:fieldequations}

We start with a brief review of modifications of general relativity displaying second-order field equations in spherical symmetry, following~\cite{Carballo-Rubio:2025ntd,Boyanov:2025pes} (see~\cite{Ziprick:2010vb,Kunstatter:2015vxa} for previous relevant work and~\cite{BorissovaEffectiveGeometrodynamics2026,BorissovaRegularBlack2026a,BorissovaAll2D2026,BorissovaModifiedFriedmann2026,Arrechea:2026ngi,Borissova$g_ttg_rr1$2026,Carballo-Rubio:2026mvj,MazzaBlackBounces2026,BorissovaFormationExtremal2026,Thaalba:2026abz} for follow-ups).
In a nutshell, any classical solutions of gravitational theories in which the field equations contain up to second-order derivatives of the metric on spherically symmetric spacetimes are captured by a 2-dimensional Horndeski theory whose degrees of freedom $g_{ab}(x)$ and $r(x)$ can be packaged into a 4-dimensional metric \qu{reconstructed} as
\begin{equation}
g_{\mu\nu}(y)\text{d}y^\mu\text{d}y^\nu=g_{ab}(x)\text{d}x^a\text{d}x^b+r^2(x)\gamma_{ij}\text{d}\theta^i\text{d}\theta^j \,;    
\end{equation}
here $\gamma_{ij}$ is the metric on the unit 2-sphere and 4-dimensional coordinates are given by $\{y^0,y^1,y^2,y^3\}=\{x^0,x^1,\theta^2,\theta^3\}$.

We will not write explicitly the 2-dimensional Horndeski Lagrangian or write the equations of motion, but rather state the result that the latter can be written as
\begin{equation}\label{eq:mfeqs}
\mathscr{G}_{\mu\nu}(q,r)=8\pi T_{\mu\nu}, 
\end{equation}
with $T_{\mu\nu}$ the stress-energy tensor of the minimally coupled matter fields and $\mathscr{G}_{\mu\nu}$ an identically conserved tensor that depends on two functions $\bm{\alpha}(r,X)$ and $\bm{\beta}(r,X)$, where we have defined $X=-\partial_a r\partial^a r/2$, with general relativity being recovered as (note that the convention in the latter definition differs by a multiplicative constant from the one in~\cite{Carballo-Rubio:2025ntd}):
\be\label{eq:alphabetagr}
\bm{\alpha} = \bm{\alpha}_\text{GR} = 2 ( 1 + 2X) \, ,
\quad
\bm{\beta}=\bm{\beta}_{\rm GR} = -2r.
\ee
Detailed expressions are not needed at this point in the discussion, and they will be written below as particular cases of the mathematical treatment to be presented below, based on biscalar 2-dimensional Horndeski theory.

The similarity with the spherically symmetric Einstein field equations does not end here. Birkhoff's theorem is generally satisfied and, for a choice of $\bm{\alpha}(r,X)$ and $\bm{\beta}(r,X)$, there is a unique static solution to Eqs.~\eqref{eq:mfeqs}~\cite{Louis-Martinez:1993agn,Ziprick:2010vb,Kunstatter:2015vxa,Carballo-Rubio:2025ntd}. Depending on the specific form of $\bm{\alpha}(r,X)$ and $\bm{\beta}(r,X)$, which must satisfy specific conditions for Eq.~\eqref{eq:mfeqs} to be meaningfully defined as a four-dimensional set of field equations~\cite{Carballo-Rubio:2025ntd,Arrechea:2026ngi,Thaalba:2026abz}, the corresponding vacuum spacetimes are either singular or regular. In general, regularity requires the introduction of a new length scale $\ell$ which is the same for all solutions of the theory and is therefore characteristic of the specific gravitational theory considered. 
For example, the Hayward metric~\cite{Hayward:2005gi} is a solution of the theory specified by
\be\label{eq:alphabetabar}
\bm{\alpha} =  \bm{\alpha}_\text{H} = \frac{2\left[r^4-3\ell^2r^2\left(1+2X\right)\right]\left(1+2X\right)}{\left[r^2-\ell^2\left(1+2X\right)\right]^2} \, ,
\quad
\bm{\beta} = \bm{\beta}_\text{H} = -\frac{2r^5}{\left[r^2-\ell^2\left(1+2X\right)\right]^2} 
\ee
--- see~\cite{Carballo-Rubio:2025ntd,Boyanov:2025pes} for further discussions.

Although we have not described how to reconstruct the Lagrangian that gives rise to these functions, it should be clear that the regularization scale $\ell$ plays the role of coupling constant, and therefore it cannot change in any dynamical process. However, if $\ell$ could be promoted to a field, then it would be allowed to vary, in principle. This can be implemented in different ways. Here we will study the simplest implementation in which $\ell$ is promoted to a scalar field --- a possibility that, to the best of our knowledge, has not been discussed before. This naturally leads to the discussion of biscalar 2-dimensional Horndeski theory below (from the perspective of the corresponding covariant gravitational actions~\cite{BorissovaRegularBlack2026a,BorissovaAll2D2026}, these theories would correspond to scalar--tensor theories with highly convoluted couplings).

There could be other ways in which a quantity $\ell$ taking constant values on solutions arises in other formalisms. For instance, it may be an integration constant (alternatively, \qu{charge} or \qu{primary hair}) associated with additional fields. An example of such a framework has been presented in~\cite{Eichhorn:2025pgy}, for a gravitational theory coupled to vector fields without dynamics (that is, without proper kinetic terms). Our approach is different in spirit, although there may be similarities worth studying.

\subsection{Biscalar 2-dimensional Horndeski}

The field equations~\eqref{eq:mfeqs} are the outcome of an action principle in which the gravitational part of the action is the 2-dimensional Horndeski action, which can be written in the so-called kinetic gravity braiding form by integrating by parts the term proportional to the 2-dimensional Ricci scalar~\cite{Takahashi:2018yzc,Colleaux:2019ckh}. This form of the 2-dimensional Horndeski action has been generalized to the biscalar case in~\cite{Nejati:2024tuo}.\footnote{The authors of~\cite{Nejati:2024tuo} proceeded as follows: they performed a spherically symmetric reduction of the usual 4-dimensional Horndeski action, thereby obtaining a particular subset of the action in \cref{eq:bilag}; then, they noted that the result could be generalized without raising the order of the equations of motion. To our knowledge, there is no proof that the action thus obtained is the most general biscalar action with second-order equations of motion in 2 dimensions. Further note that, technically, such a formalism is suitable for any number of fields.} 
The resulting biscalar Lagrangian, which we will be using throughout the paper, is the following:
\be \label{eq:bilag}
\mathcal{L} = F \left( \phi^I, X_{JK} \right) + G^I \left( \phi^J, X_{KL} \right) \Box \phi^I \, ,
\ee
where summation over repeated indices is assumed, and we have defined the variable 
\be
X_{IJ} := - \frac{1}{2} \nabla_a \phi^I \nabla^a \phi^J \, ,
\ee
which is symmetric in its indices.
For the Lagrangian in Eq.~\eqref{eq:bilag} to have second-order field equations, and therefore represent a genuine biscalar generalization of the 2-dimensional Horndeski action, the following relations must be satisfied (see the proof of this statement in App.~\ref{app:tseqs}):
\be \label{eq:grel2nd}
G^K_{\ X_{IJ}} = G^I_{\ X_{KJ}} \, ,
\ee
meaning that $G^K_{\ X_{IJ}}$ must be completely symmetric in its indices. In the equation above, we have introduced for the first time a notation that we will be using heavily below, in which a variable appearing as subindex indicates the derivative with respect to that variable, e.g.~$G^K_{\ X_{IJ}}=\partial G^K/\partial{X_{IJ}}$.

Under these conditions, the field equations that result from variations of the fields $g_{ab}$ and $\phi^I$ are, respectively (see~\cref{app:tseqs} for a complete derivation):
\be
\label{eq:tensor}
\mathcal{E}_{ab} &=  2 X_{IJ}G^I_{\ X_{JK} }\left( \nabla_a\nabla_b\phi^K-g_{ab}\square\phi^K \right)
+g_{ab}\left( F + 2 G^I_{\ \phi^J} X_{IJ} \right)\nonumber \\
&\phantom{=} +  \left(F_{X_{IJ}}+ 2 G^I_{\ \phi^J} \right)\nabla_{(a} \phi^I \nabla_{b)} \phi^J=0\, ,\\
\label{eq:scalar}
\mathcal{E}^I &= F_{\phi^I} - 2 X_{JK} \left( F_{X_{IJ}\phi^K} + G^I_{\ \phi^J \phi^K}  \right)
+ \left(F_{X_{IJ}} + G^J_{\ \phi^I} + G^I_{\ \phi^J} - 2 X_{LK}G^I_{\ \phi^L X_{KJ}}  \right) \Box \phi^J \nonumber\\
&\phantom{=} + \left(F_{X_{IJ}X_{KL}}  +2 G^I_{\ \phi^J X_{KL}}\right)\nabla^a \phi^J \nabla_a X_{KL}
+ X_{JK} G^K_{\ X_{IJ}} R 
\nonumber\\
&\phantom{=}
+ \left(G^I_{\ X_{JK}} +X_{LM} G^I_{\ X_{JK} X_{LM}}\right)\left( \Box \phi^J \Box \phi^K
- \nabla^a \nabla^b \phi^J \nabla_a \nabla_b \phi^K \right)=0
\, .
\ee
These equations are not completely independent, as they are related by the following Bianchi identity:
\be\label{eq:Bianchi}
\nabla^a \mathcal{E}_{ab} + \frac{1}{2} \mathcal{E}^I \nabla_b \phi^I = 0 \, .
\ee

To make contact with the notation and works reviewed in the previous section, we may introduce the following quantities:
\be
\bm{\beta}^I &:= 2 X_{JK} G^I_{\ X_{JK}} \label{eq:defbeta}
\, ,\\
- \frac{1}{2} \bm{\alpha} &:= F + X_{IJ} \left( G^I_{\ \phi^J} + G^J_{\ \phi^I} \right) \label{eq:defalpha}
\, ,\\
\bm{\gamma}^{IJ} &:= F_{X_{IJ}} + G^I_{\ \phi^J} + G^J_{\ \phi^I} 
\label{eq:defgamma}
\, .
\ee
Note that
\be
\label{eq:gammaID}
\bm{\gamma}^{IJ} &= - \frac{1}{2} \bm{\alpha}_{X_{IJ}} - \frac{1}{2} \left( \bm{\beta}^I_{\ \phi^J} + \bm{\beta}^J_{\ \phi^I} \right) + X_{KL} \left( \bm{\gamma}^{IJ}_{\ X_{KL}} - \bm{\gamma}^{KL}_{\ X_{IJ}} \right) \, ,
\\
\label{eq:betaID}
\bm{\beta}^I_{\ \phi^J} - \bm{\beta}^J_{\ \phi^I} &= - 2 X_{KL} \left( \bm{\gamma}^{IK}_{\ X_{JL}} - \bm{\gamma}^{JL}_{\ X_{IK}} \right) \, .
\ee
The equations of motion then become
\be
\label{eq:tensor_alphabeta}
\mathcal{E}_{ab} &=  \bm{\beta}^I\left( \nabla_a\nabla_b\phi^I-g_{ab}\square\phi^I \right)
-\frac{1}{2}\bm{\alpha} g_{ab}+\bm{\gamma}^{IJ}\nabla_{(a} \phi^I \nabla_{b)} \phi^J=0\, ,\\
\label{eq:scalar_alphabeta}
\mathcal{E}^I &= -\frac{1}{2}\bm{\alpha}_{\phi^I} - 2 X_{JK} \bm{\gamma}^{IJ}_{\ \phi^K}
+ \left(\bm{\gamma}^{IJ} - 2 X_{LK} G^I_{\ \phi^L X_{JK}}  \right) \Box \phi^J 
\nonumber\\
&\phantom{=}  
+ \left(\bm{\gamma}^{IJ}_{\ X_{KL}}  + G^I_{\ \phi^J X_{KL}} - G^J_{\ \phi^I X_{KL}}\right)\nabla^a \phi^J \nabla_a X_{KL}
+ \frac{1}{2}\bm{\beta}^I R
\nonumber\\
&\phantom{=}
+ \frac{1}{2}\bm{\beta}^I_{\ X_{JK}}\left( \Box \phi^J \Box \phi^K
- \nabla^a \nabla^b \phi^J \nabla_a \nabla_b \phi^K \right)
=0 \, .
\ee
In the following, we shall employ either notations, depending on which offers a clearer picture.

It is important to stress that reconstructing the functions $F$ and $G^I$ in Eq.~\eqref{eq:bilag} in terms of $\bm{\alpha}$, $\bm{\beta}^I$, and $\bm{\gamma}^{IJ}$ is not trivial and, generally, cannot be done in a unique way. Further note that, while $\bm{\alpha}$, $\bm{\beta}^I$, and $\bm{\gamma}^{IJ}$ suffice to write the tensorial equations, the equations for the scalar fields require additional information encoded in the derivatives of $G^I$.

\subsection{Unitary gauge} 

The equations of motion of the biscalar Horndeski theory take a simpler form when expressed in the particular gauge in which one of the two scalar fields, say $\phi^1$, is used as a coordinate of the 2-dimensional spacetime.\footnote{The conditions $\partial\phi^1/\partial x^\mu\neq0$ are sufficient to guarantee that this is always possible. More generally, one can use both fields as coordinates as long as $\det(\nabla_\mu \phi^1, \nabla_\nu \phi^2) \neq 0$; i.e., if $\phi^1\sim r$, as long as $\pdv*{\phi}{v}\neq 0$, one can take $\phi^2 \sim v$. This gauge is reminiscent of the unitary gauge in e.g.~cosmological \cite{PiazzaEffectiveField2013} and black hole perturbation theory \cite{FrancioliniEffectiveField2019,MukohyamaEffectiveField2025}.} 
This choice is especially motivated if the 2-dimensional Horndeski theory is thought of as the spherically symmetric reduction of some 4-dimensional theory --- coherently with the derivation of \cite{Nejati:2024tuo} and with the spirit of \cite{Carballo-Rubio:2025ntd}. In such a case, one of the two scalars is naturally interpreted as a dilaton emerging from the dimensional reduction, while the other is a genuinely new degree of freedom.

We will call $v$ and $r$ the spacetime coordinates, and write the metric as
\be\label{eq:metric}
g_{ab} \dd{x^a} \dd{x^b} = - f(v,r) \dd{v^2} + 2 h(v,r) \dd{v} \dd{r} \, .
\ee
For simplicity, we will focus on the particular case $h(v,r) = 1$: although this choice \emph{does} entail a loss of generality~\cite{MazzaBlackBounces2026}, we deem it appropriate for the scope of this article.
We will then identify $\phi^1$ with $r$, i.e.~we assume that coordinates are chosen so that
\be
\pdv{\phi^1}{v} = 0
\qq{and}
\pdv{\phi^1}{r} = 1 \, .
\ee
For clarity, however, we will mostly preserve the distinction between the derivatives with respect to the field $\phi^1$ and the spacetime coordinate $r$ --- that is, for example, the derivative $\pdv*{\phi^2}{r}$ is clearly \emph{not equal} to $\pdv*{\phi^2}{\phi^1}$.

In this gauge, the three independent components of the tensor $\mathcal{E}_{ab}$ occurring in the tensorial equations $\mathcal{E}_{ab}=0$ read
\bse\label{eq:tensor_gauge}
\be
\label{eq:Err_gauge}
\mathcal{E}_{rr} &= \bm{\gamma}^{11} + \bm{\beta}^2 \pdv[2]{\phi^{2}}{r} + 2 \bm{\gamma}^{12} \pdv{\phi^{2}}{r} + \bm{\gamma}^{22} \pdv{\phi^{2}}{v} \pdv{\phi^{2}}{r} \, , \\
\label{eq:Evr_gauge}
\mathcal{E}_{vr} &= - \frac{1}{2} \bm{\alpha} - \frac{1}{2} \bm{\beta}^1 \pdv{f}{r} - \bm{\beta}^2 \left( \frac{1}{2} \pdv{f}{r} \pdv{\phi^{2}}{r} + f \pdv[2]{\phi^{2}}{r} + \pdv{\phi^{2}}{v}{r}  \right) \nonumber\\
&\phantom{=} + \bm{\gamma}^{12} \pdv{\phi^{2}}{v} + \bm{\gamma}^{22} \pdv{\phi^{2}}{v} \pdv{\phi^{2}}{r} \, ,\\
\label{eq:Evv_gauge}
 \mathcal{E}_{vv} + f\mathcal{E}_{vr}  &= \frac{1}{2} \bm{\beta}^1 \pdv{f}{v} + \frac{1}{2} \bm{\beta}^2 \left( \pdv{f}{v} \pdv{\phi^{2}}{r} - \pdv{f}{r} \pdv{\phi^{2}}{v} + 2 f \pdv{\phi^{2}}{v}{r} + 2 \pdv[2]{\phi^{2}}{v} \right) \nonumber\\
&\phantom{=} + \left[\bm{\gamma}^{12} f + \bm{\gamma}^{22} \left( f \pdv{ \phi^{2}}{r} + \pdv{\phi^{2}}{v} \right) \right] \pdv{\phi^{2}}{v} \, .
\ee\ese
The scalar equations are much more involved and not particularly insightful at this point, hence we omit them for the time being.
Moreover, the Bianchi identity of \cref{eq:Bianchi} ensures that
\be\label{eq:Bianghi_gauge}
\mathcal{E}_{ab} = 0 
\Rightarrow
\begin{cases}
\mathcal{E}^2 \pdv{\phi^{2}}{v} = 0 \, ,\\
\mathcal{E}^1 + \mathcal{E}^2 \pdv{\phi^{2}}{r} = 0 \, .
\end{cases} 
\ee
Hence, for configurations such that $\pdv*{\phi^{2}}{v} \neq 0$, the tensorial equations imply both scalar equations; if instead $\phi^{2}$ does not depend on time, the scalar equation relative to $\phi^{2}$ itself must be solved independently. In both cases, the remaining scalar equation $\mathcal{E}^1 = 0$ follows as a consequence of the others.

For future reference, we also report the explicit form of the kinetic terms $X_{IJ}$, expressed in this gauge:
\bse\label{eq:XIJgauge}
\be
X_{11} &= - \frac{1}{2} f \, ,\\
X_{12} &= - \frac{1}{2} \left( \pdv{\phi^2}{v} + f \pdv{\phi^2}{r} \right) \, ,\\
X_{22} &= - \frac{1}{2} \pdv{\phi^2}{r} \left( \pdv{\phi^2}{v} + 2 f \pdv{\phi^2}{r} \right) \, .
\ee
\ese

\section{The single-field case and its extension\label{sec:singext}}

The framework described in the previous section encompasses the one developed in~\cite{Carballo-Rubio:2025ntd,Boyanov:2025pes} in a rather natural way. Indeed, the theories considered in those references and in the follow-up works~\cite{BorissovaEffectiveGeometrodynamics2026,BorissovaRegularBlack2026a,BorissovaAll2D2026,BorissovaModifiedFriedmann2026,Arrechea:2026ngi,Borissova$g_ttg_rr1$2026,Carballo-Rubio:2026mvj,MazzaBlackBounces2026,BorissovaFormationExtremal2026,Thaalba:2026abz} can be understood as particular biscalar Horndeski theories in which $G^2 = 0$, while $F(\phi^1,X_{11})$ and $G^1(\phi^1,X_{11})$ depend only on $\phi^1$ and $X_{11}$, which we will refer to as the \emph{single-field case}.

Given a particular single-field theory, therefore, extending it to a biscalar theory requires promoting the two functions $F(\phi^1,X_{11})$ and $G^1(\phi^1,X_{11})$ to $F(\phi^I,X_{JK})$ and $G^1(\phi^I,X_{JK})$, as well as introducing a new function $G^2(\phi^I,X_{JK})$. The simplest way of implementing such an extension appears to be upgrading the regularization scale $\ell$ to a function $\ell(\phi^2, X_{12}, X_{22})$ of the additional dynamical variables. Ideally, in this way one should be able to recover previous results in specific situations (e.g.~imposing that $\phi^2$ be independent of time, or be a constant); at the same time, it becomes very natural to envisage the possibility that $\ell$ could change in time, in a way that would be dictated by the field equations, depending on the specific functional dependence $\ell(\phi^2, X_{12}, X_{22})$ considered.

Explicitly, this strategy entails promoting the functions in the single-field Lagrangian so that the parametric dependence on $\ell$ becomes a functional dependence on $\phi^2$, $X_{12}$ and $X_{22}$. This extension is not unique, and therefore we need to explore different ways in which this procedure can be implemented.

First of all, it is useful to note that imposing $h(v,r)=1$ in Eq.~\eqref{eq:metric} implies, in the single-field case, that there exists a \qu{potential} function $\bm{\Omega}$, defined up to a multiplicative constant by the condition $-\bm{\alpha}_{X_{11}}/2 - \bm{\beta}^1_{\ \phi^1} = 0$, such that
\bse
\be
\bm{\alpha} &= \bm{\Omega}_{\phi^1} \left(\phi^1, X_{11}, \ell \right) \, , \\
-2\bm{\beta}^1 &= \bm{\Omega}_{X_{11}} \left(\phi^1, X_{11}, \ell \right) \, ;
\ee\ese
see the explicit discussion in~\cite{Carballo-Rubio:2025ntd,Boyanov:2025pes}.
This suggests the specific form of these functions in the biscalar case:
\bse\label{eq:naive_alphabeta}
\be
\bm{\alpha} &= \bm{\Omega}_{\phi^1} \left(\phi^1, X_{11}, \ell(\phi^2, X_{12}, X_{22}) \right) \, , \\
-2\bm{\beta}^1 &= \bm{\Omega}_{X_{11}} \left(\phi^1, X_{11}, \ell(\phi^2, X_{12}, X_{22}) \right) \, ; 
\ee\ese 
i.e.~$\bm{\alpha}$ and $\bm{\beta}^1$ are assumed to have the same functional form as in the single-field case, except for their dependence on $\phi^2$, $X_{12}$, and $X_{22}$, which however appears only through $\ell$. Even if we fix these functions as stated above, there is still residual freedom as one has to specify the functions $\bm{\beta}^2$ and $\bm{\gamma}^{IJ}$.

We first start considering the possibility that
\be\label{eq:noG2}
G^2 = 0 \, .
\ee
This entails that the Lagrangian is extended in a \qu{minimal} way, in the sense that all novelty with respect to the single-field case is encoded in the dependence of $F$ and $G^1$ on $\phi^2$, $X_{12}$, and $X_{22}$. The symmetry of the $G^I_{\ X_{JK}}$, encoded in \cref{eq:grel2nd}, then implies $G^1_{\ X_{12}} = 0 = G^1_{\ X_{22}}$.
Hence, some immediate consequences of \cref{eq:noG2} are that
\be
\bm{\beta}^2 = 0\, ,
\ee
and
\be
\bm{\beta}^1 = 2X_{11} G^1_{\ X_{11}} \, ,
\qq{with} 
\bm{\beta}^1_{\ X_{12}} = 0 = \bm{\beta}^1_{\ X_{22}} \, .
\ee
Via \cref{eq:naive_alphabeta}, this entails that
\be
\bm{\Omega}_{\ell X_{11}} \pdv{\ell}{X_{12}} = 0 = \bm{\Omega}_{\ell X_{11}} \pdv{\ell}{X_{22}} \, ,
\ee
and since we have assumed $\Omega_{\ell X_{11}} \neq 0$ we deduce that
\be
\pdv{\ell}{X_{12}} = 0
\qq{and}
\pdv{\ell}{X_{22}} = 0 \, .
\ee
Moreover, according to \cref{eq:defgamma}, \cref{eq:noG2} implies
\be
\bm{\gamma}^{11} &= - \frac{1}{2} \bm{\alpha}_{X_{11}} - \bm{\beta}^1_{\ \phi^1} - X_{12} G^1_{\ \phi^2 X_{11}}\, ,\\
\bm{\gamma}^{12} &=- \frac{1}{2} \bm{\alpha}_{X_{12}} \, \\
\bm{\gamma}^{22} &= - \frac{1}{2} \bm{\alpha}_{X_{22}} \, ;
\ee
hence, via \cref{eq:naive_alphabeta} and the results above, we automatically have
\be
\bm{\gamma}^{12} &= \bm{\gamma}^{22} = 0 \, ,\\
\bm{\gamma}^{11} &= - X_{12} G^1_{\ \phi^2 X_{11}} \nonumber\\
&= - \frac{1}{2} \frac{X_{12}}{X_{11}} \bm{\Omega}_{\ell X_{11}} \pdv{\ell}{\phi^2} \, .
\ee
Note that, since $\bm{\beta}^2$, $\bm{\gamma}^{12}$, and $\bm{\gamma}^{22}$ all vanish, the equation of motion $\mathcal{E}_{rr} = 0$ forces $\bm{\gamma}^{11}$ to vanish as well, at least on the solutions to the equations of motion.

The \qu{minimal} assumption of \cref{eq:noG2} thus turns out to be extremely constraining, since it completely rules out the possibility that $\ell$ depend on $X_{12}$ and $X_{22}$; moreover, the equations of motion imply that $\ell$ cannot depend on $\phi^2$ either, \emph{at least on all configurations such that} $X_{12} \neq 0$.
Leaving this caveat aside for the moment, the conclusion seems rather clear-cut: in general, a consistent biscalar extension of the single-field case appears to necessitate the introduction of a non-vanishing $G^2$. 

It is worth noting, however, that the caveat mentioned above leaves some wiggle room: on configurations such that $X_{12} = 0$, the conclusion that, in the minimal extension implemented by \cref{eq:noG2}, $\ell$ must be independent from $\phi^2$ does not hold. 
Confronting with \cref{eq:XIJgauge}, we realize that $X_{12} = 0$ if $\phi^2 = \text{const.}$, for instance. 

In the next section, we thus choose to analyze the $\phi^2 = \text{const.}$ particular case. However, we will do so without assuming $G^2 = 0$ from the start, with the intention of specifying the general results when needed. Hence, this choice, while suggested by the \qu{minimal} assumption just described, is fairly independent from the specific way in which the single-field case in extended; rather, it represents a branching point in our exploration of the subject. It is understood that other choices ought to be investigated as well.

Moreover, we will devote particular attention to solutions describing a static metric. Note that this does not contradict the motivation of the whole paper, namely the development of a framework whereby $\ell$ is allowed to vary, since, in our approach, static solutions represent the different equilibrium states that dynamical transients connect. Hence, the existence of static solutions actually constitutes a basic requirement of the framework we wish to develop, and is thus the first necessary ingredient for the whole construction to make sense.

\section{Constant scalar\label{sec:const_scalar}}

As a first step, we shall consider the case in which $\phi^2$ is constant. As mentioned above, and as we will show below, this is coherent with the examples discussed above; moreover, it entails substantial simplifications. 

\subsection{Field equations}

We take $\phi^2=\text{const.}$, i.e.
\be\label{eq:constant_phi}
\pdv{\phi^{2}}{v} = 0 = \pdv{\phi^{2}}{r} \, .
\ee
Confronting with \cref{eq:XIJgauge}, we notice that this implies
\be
X_{12} = 0 = X_{22} \, ,
\ee
and the only new variable, with respect to the single-field case, is therefore the constant value of $\phi^2$ itself. The (constant) regularization scale $\ell$ must therefore depend on $\phi^2$, coherently with the discussion in the section above.
Moreover, from Eq.~\eqref{eq:gammaID} we have the following simplification:
\be
\bm{\gamma}^{11} = - \frac{1}{2} \bm{\alpha}_{X_{11}} - \bm{\beta}^1_{\ \phi^1}
\, ,
\ee
and this vanishes identically for the naive extensions of the single-field case we are considering --- cf.~\cref{eq:naive_alphabeta}.

Confronting with the third tensorial equation, Eq.~\eqref{eq:Evv_gauge}, we immediately realize that the assumptions of \cref{eq:constant_phi} imply
\be\label{eq:static_metric}
\pdv{f}{v} = 0 \, ,
\ee
hence all solutions with $\phi^2=\text{const.}$ are necessarily static.
The remaining tensorial Eqs.~\eqref{eq:Err_gauge} and~\eqref{eq:Evr_gauge} reduce to
\bse\label{eq:tensor_gauge_static}
\be
\label{eq:Err_static}
\mathcal{E}_{rr} &= \bm{\gamma}^{11} =0 \, , \\
\label{eq:Evr_static}
\mathcal{E}_{vr} &= - \frac{1}{2} \bm{\alpha} - \frac{1}{2} \bm{\beta}^1 \pdv{f}{r}=0  \, ,
\ee
\ese
which are identically satisfied by construction for the direct extensions of~\cref{eq:naive_alphabeta}.
By virtue of the Bianchi identity of \cref{eq:Bianghi_gauge}, the field equation for $\phi^1$ is automatically satisfied when the tensorial equations are satisfied.
The field equation for $\phi^2$ instead reads
\be
\label{eq:E2_static}
0 = \mathcal{E}^2 &= F_{\phi^2} -2 X_{11} \left( F_{X_{12}\phi^1} + G^2_{\ \phi^1 \phi^1} \right) +  \left(F_{X_{12}} + G^1_{\ \phi^2} +  G^2_{\ \phi^1}- 2 X_{11} G^2_{\ \phi^1 X_{11}}  \right) \Box \phi^1 \nonumber\\
&\phantom{=}+ \left(F_{X_{12}X_{11}}  +2 G^2_{\ \phi^1 X_{11}}\right)\nabla^a \phi^1 \nabla_a X_{11}\nonumber\\
&\phantom{=}
+ X_{11} G^2_{\ X_{11}} R
+ \left(G^2_{\ X_{11}} +X_{11} G^2_{\ X_{11} X_{11}}\right)\left( \Box \phi^1 \Box \phi^1
- \nabla^a \nabla^b \phi^1 \nabla_a \nabla_b \phi^1 \right)\\
\label{eq:E2_static_alphabeta}
&= -\frac{1}{2}\bm{\alpha}_{\phi^2} - 2 X_{11} \bm{\gamma}^{12}_{\ \phi^1}
+ \left(\bm{\gamma}^{12} - \bm{\beta}^2_{\ \phi^1} \right) \Box \phi^1 \nonumber\\
&\phantom{=} +\left(\bm{\gamma}^{12}_{\ X_{11}}  + G^2_{\ \phi^1 X_{11}} - G^1_{\ \phi^2 X_{11}}\right)\nabla^a \phi^1 \nabla_a X_{11}
\nonumber\\
&\phantom{=} + \frac{1}{2}\bm{\beta}^2 R + \frac{1}{2}\bm{\beta}^2_{\ X_{11}}\left( \Box \phi^1 \Box \phi^1
- \nabla^a \nabla^b \phi^1 \nabla_a \nabla_b \phi^1 \right)
\, ,
\ee
and this is now an independent equation, since the Bianchi identity does not connect it to the tensorial equations if $\pdv*{\phi^{2}}{v} = 0$.
The conditions of \cref{eq:constant_phi,eq:static_metric} entail some rather drastic simplifications, including the following identities: 

\bse\label{eq:static_identities}
\noindent\centering
\begin{minipage}{0.35\textwidth}
\be
\Box \phi^1 = \pdv{f}{r} \, &,\\
R = - \pdv[2]{f}{r} \, &,
\ee
\end{minipage}
\hfill
\begin{minipage}{0.6\textwidth}
\be
\Box \phi^1 \Box \phi^1 - \nabla^a \nabla^b \phi^1 \nabla_a \nabla_b \phi^1 = \frac{1}{2} \left( \pdv{f}{r} \right)^2 \, &,\\
\nabla^a \phi^1 \nabla_a X_{11} = X_{11} \Box \phi^1 \, &, \label{eq:static_identities_useful}
\ee
\end{minipage}
\ese

\bigskip

\noindent Using \cref{eq:static_identities_useful}, the scalar field's equation simplifies to
\be\label{eq:e2redef}
\mathcal{E}^2 &= F_{\phi^2} - 2 X_{11}\left( F_{X_{12}\phi^1} + G^2_{\ \phi^1 \phi^1} \right) +  \left( F_{X_{12}} + X_{11} F_{X_{11} X_{12}} + G^1_{\ \phi^2} +  G^2_{\ \phi^1}  \right) \Box \phi^1 \nonumber\\
&\phantom{=}
+ X_{11} G^2_{\ X_{11}} R
+ \left(X_{11} G^2_{\ X_{11}}\right)_{X_{11}}\left( \Box \phi^1 \Box \phi^1
- \nabla^a \nabla^b \phi^1 \nabla_a \nabla_b \phi^1 \right) \\
&= -\frac{1}{2} \bm{\alpha}_{\phi^2} - 2 X_{11} \bm{\gamma}^{12}_{\ \phi^1}
+\left[ \left(X_{11} \bm{\gamma}^{12} \right)_{X_{11}} - \frac{1}{2} \bm{\beta}^1_{\ \phi^2} - \frac{1}{2} \bm{\beta}^2_{\ \phi^1} \right] \Box \phi^1 \nonumber\\
&\phantom{=} + \frac{1}{2} \left[ \bm{\beta}^2 R + \bm{\beta}^2_{\ X_{11}} \left( \Box \phi^1 \Box \phi^1
- \nabla^a \nabla^b \phi^1 \nabla_a \nabla_b \phi^1 \right) \right]\, .
\ee

An important remark is in order at this point. For a constant value of $\phi^2$ and a static metric, there are three independent equations of motion --- $\mathcal{E}_{rr}=0$, $\mathcal{E}_{vr}=0$, and $\mathcal{E}^{2}=0$ --- for only one unknown --- the metric function $f(r)$. (The constant value of $\phi^2$ could also be considered unknown, and to be determined via the equations of motion, but this is irrelevant for what follows.)
Hence, some of these equations must be interpreted as constraints on the functional form of the Lagrangian. 

Notably, however, there is no unique prescription to derive constraints on the Lagrangian from an on-shell statement such as, for example, $\mathcal{E}^2=0$. Another branching point in the treatment, therefore, is associated with the decision of imposing these constraints on shell, i.e.~as relations that are only valid when the equations of motion are satisfied; or off shell, i.e.~valid for any field configuration. Clearly, an off-shell constraint is also valid on shell, and therefore it is a stronger statement. 

In the following, we shall consider imposing off-shell constraints. We will start with stronger sets of constrains that are too stringent for our scope, and subsequently relax them to more general statements.

\subsection{Identifying off-shell constraints\label{sec:offshell}}

We define the off-shell approach as the identification of a series of off-shell constraints for the coupling functions in the Lagrangian density~\eqref{eq:bilag} that are equivalent to the equation of motion $\mathcal{E}^2=0$. The latter equation contains however a mix of the couplings functions and differential operators acting on the field $\phi^1$ or the metric $g_{ab}$, which means that this procedure is not unique.

To make this explicit, let us start considering the equation of motion $\mathcal{E}^2=0$ in its original form in Eq.~\eqref{eq:E2_static}. Aside from a term that does not contain derivatives of $\phi^1$ explicitly, we can identify the scalar combinations $\square\phi^1$, $\nabla^a\phi^1\nabla_aX_{11}$, $R$ and $\square\phi^1\square\phi^1-\nabla^a\nabla^b\phi^1\nabla_a\nabla_b\phi^1$, which are multiplied by combinations of the coupling functions and their derivatives with respect to their arguments. By demanding that all these combinations vanish off shell, we could guarantee the validity of the on-shell equation of motion $\mathcal{E}^2=0$. In this case, we would have a total of 5 (a priori independent) relations. The implications of this choice are actually equivalent to the case discussed below, arising for a different choice that results when taking into account on-shell degeneracies in some of the terms in the above equation of motion, and it is thus implicitly contained in the subsequent discussion.

We have already discussed some on-shell relations in Eq.~\eqref{eq:static_identities} that allowed us to simplify the form of the equation of motion $\mathcal{E}^2=0$, leading to \cref{eq:e2redef}. Requiring that each of the functions that multiply the different independent scalars built from $\phi^1$ and the metric and their derivatives in \cref{eq:e2redef} vanish identically now translates into 3 relations:
\be
F_{\phi^2} - 2X_{11}\left(F_{X_{12}\phi^1} +G^2_{\ \phi^1 \phi^1}\right) &= 0\, ,\label{eq:firstrel}\\
F_{X_{12}} + X_{11} F_{X_{11} X_{12}} + G^1_{\ \phi^2} +  G^2_{\ \phi^1} &= 0\, ,\label{eq:secondrel}\\
X_{11} G^2_{\ X_{11}} &= 0\, \label{eq:thirdrel}.
\ee
Note that these relations are completely independent from \cref{eq:naive_alphabeta}: they do not follow directly from them, nor do they appear to contradict them.
Moreover, the last of these relations entails that $\bm{\beta}^2=0$ (at least on configurations for which $\phi^2=\text{const.}$), and is therefore closely related to \cref{eq:noG2}.

It is thus worth comparing in further detail these constraints with those that arise from the simple extension of the single-field case encoded in \cref{eq:naive_alphabeta}, along with the minimality assumption of \cref{eq:noG2}. First of all, note that \cref{eq:thirdrel} implies $G^2_{\ X_{11}}=0$ and, due to the symmetry of $G^I_{\ X_{JK}}$, also that $G^1_{\ X_{12}}=0$.
Recalling that, for $\phi^2 = \text{const.}$, we have $\bm{\beta}^1=2X_{11}G^1_{\ X_{11}}$, taking its derivative with respect to $X_{12}$ therefore yields 
\be
\bm{\beta}^1_{\ X_{12}}=0\, .
\ee
According to \cref{eq:naive_alphabeta}, this already entails 
\be
\pdv{\ell}{X_{12}} = 0\,.
\ee
This result is analogous to the one derived in \cref{sec:singext}, but contrary to that section here we do not assume $G^2 = 0$ [i.e.~\cref{eq:noG2}] from the outset. Furthermore, this result entails that the whole Lagrangian, including the function $F$, cannot depend on $X_{12}$. Using this fact, taking a derivative with respect to $X_{11}$ of the second relation~\eqref{eq:secondrel}, and using the derivative with respect to $\phi^1$ of the third relation~\eqref{eq:thirdrel}, we thus get
\be
G^1_{\ \phi^2 X_{11}} = 0 \, ,
\ee
but this implies
\be
\bm{\beta}^1_{\ \phi^2} = 0 \Rightarrow \pdv{\ell}{\phi^2} = 0\, .
\ee
This result too is analogous to the one derived in \cref{sec:singext}, and does not require \cref{eq:noG2}.
We deduce that the single-field case cannot be extended in this way, namely implementing Eqs.~(\ref{eq:firstrel}-\ref{eq:thirdrel}) together with Eqs.~(\ref{eq:naive_alphabeta}).

Hence, this identification of the constraints in the off-shell approach to the case $\phi^2 = \text{const.}$ does not seem to lead to a viable construction when the relations in Eqs.~(\ref{eq:naive_alphabeta}), widely used in the single-field case, are simultaneously considered. The 3 independent relations obtained are incompatible with a solution (which implies that the original case with 5 independent off-shell relations is also incompatible). This is not particularly discouraging, however, since the identification of the off-shell constraints is not unique. Moreover, this result does not preclude the possibility that the single-field case be extended in some less naive manner than what is entailed by \cref{eq:naive_alphabeta}.

As a final step along this line of reasoning, we show how to further reduce the 3 off-shell constraints to a single one, which is the minimal case in the sense that less assumptions are needed, while the previous cases are included. This can be achieved by fully exploiting the relations in \cref{eq:static_identities}, and by using the tensorial equation of motion $\mathcal{E}_{vr} = 0$ to write the derivative of $f(\phi^1)$ as [cf.~\cref{eq:Evr_gauge}]:
\be
f'(\phi^1)=\frac{F + 2X_{11} G^1_{\ \phi^1} }{X_{11}G^1_{\ X_{11}}} \, .   
\ee
This relation can be used to write its second derivative, $f''(\phi^1)$, entirely in terms of the coupling functions and its derivatives as:
\be
f''(\phi^1)&=\frac{F_{\phi^1}+2X_{11}G^1_{\ \phi^1\phi^1}}{X_{11}G^1_{\ X_{11}}}-\frac{F+2X_{11}G^1_{\ \phi^1}}{X_{11}(G^1_{\ X_{11}})^2}G^1_{\ X_{11}\phi^1} -  \left[\frac{F_{X_{11}}+2X_{11} G^1_{\ \phi^1 X_{11}}+2G^1_{\ \phi^1}}{X_{11}G^1_{\ X_{11}}}
\right.
\nonumber\\
&\phantom{=}
\left.
-\frac{F+2X_{11}G^1_{\ \phi^1}}{(X_{11}G^1_{\ X_{11}})^2}\left(G^1_{\ X_{11}}+X_{11}G^1_{\ X_{11}X_{11}}\right)\right] \frac{F+2X_{11}G^1_{\ \phi^1}}{2X_{11}G^1_{\ X_{11}}} \, .
\ee
Inserting the expressions above for $f'(\phi^1)$ and $f''(\phi^1)$ results in the identification of the following off-shell constraint:
\be
&\phantom{=} F_{\phi^2} 
-2 X_{11} ( F_{X_{21}\phi^1} + G^2_{\ \phi^1 \phi^1})\nonumber\\
&\phantom{=}+(F_{X_{21}}+X_{11}F_{X_{12}X_{11}} + G^1_{\ \phi^2} +  G^2_{\ \phi^1}) \frac{F+2X_{11} G^1_{\ \phi^1}}{X_{11}G^1_{\ X_{11}}} \nonumber\\
&\phantom{=}
+ \frac{1}{2}\left(G^2_{\ X_{11}} +X_{11} G^2_{\ X_{11} X_{11}}\right)\left(\frac{F+2X_{11}G^1_{\ \phi^1}}{X_{11}G^1_{\ X_{11}}} \right)^2\nonumber\\
&\phantom{=}-X_{11}G^2_{\ X_{11} } \left[\frac{F_{\phi^1}+2X_{11}G^1_{\ \phi^1\phi^1}}{X_{11}G^1_{\ X_{11}}}+\frac{(F+2X_{11}G^1_{\ \phi^1})^2\left(G^1_{\ X_{11}}+X_{11}G^1_{\ X_{11}X_{11}}\right)}{2(X_{11}G^1_{\ X_{11}})^3} 
\right.\nonumber\\
&\phantom{=}\left. -\frac{2X_{11}(F+2X_{11}G^1_{\ \phi^1})G^1_{\ X_{11}\phi^1}+ (F_{X_{11}}+2X_{11}G^1_{\ \phi^1 X_{11}}+2G^1_{\ \phi^1})(F+2X_{11}G^1_{\ \phi^1})}{2 (X_{11}G^1_{\ X_{11}})^2}\right]=0\, .
\ee
This equation can be rearranged and written in a more schematic way to make clear that it is a linear partial differential equation for the function $G^2(\phi^1,\phi^2,X_{11},X_{12},X_{22})$:
\begin{equation}\label{eq:genoffc}
\mathcal{C}_1 G^2_{\phi^1\phi^1}+\mathcal{C}_2G^2_{X_{11}X_{11}}+\mathcal{C}_3G^2_{\phi^1}+\mathcal{C}_4G^2_{X_{11}}+\mathcal{S}=0\, ,
\end{equation}
where we have defined
\be
\mathcal{C}_1&:= - 4 X_{11}\left(X_{11}G^1_{\ X_{11}}\right)^3,\nonumber\\ 
\mathcal{C}_2&:= X_{11}^2 G^1_{\ X_{11}} \left(F+2X_{11}G^1_{\ \phi^1} \right)^2, \nonumber\\  
\mathcal{C}_3&:=2 \left( X_{11} G^1_{\ X_{11}} \right)^2 \left(F+2X_{11}G^1_{\ \phi^1}\right), \nonumber\\  
\mathcal{C}_4&:= X_{11} G^1_{\ X_{11}} \left(F+2X_{11}G^1_{\ \phi^1} \right)^2 - X_{11} \left\{  2\left( X_{11}G^1_{\ X_{11}} \right)^2 \left(F_{\phi^1}+2G^1_{\ \phi^1\phi^1}X_{11} \right) \right.\nonumber\\
&\phantom{=}\left. - \left(X_{11}G^1_{\ X_{11}} \right) \left[2X_{11}\left(F+2X_{11}G^1_{\ \phi^1}\right) G^1_{\ X_{11}\phi^1} + 
\left(F_{X_{11}}+2G^1_{\ \phi^1 X_{11}}X_{11}+2G^1_{\ \phi^1}\right)\left(F+2G^1_{\ \phi^1}X_{11}\right)\right]\right.\nonumber\\
&\phantom{=}\left. +\left(F+2G^1_{\ \phi^1}X_{11}\right)^2\left(G^1_{\ X_{11}}+X_{11}G^1_{\ X_{11}X_{11}}\right)\right\} \, , \nonumber\\
\mathcal{S} &:= 2 \left(X_{11}G^1_{\ X_{ 11}}\right)^3\left(F_{\phi^2} - 2X_{11} F_{\phi^1 X_{12}} \right) + 2 \left(X_{11}G^1_{\ X_{11}}\right)^2 \left( F_{X_{12}} + X_{11} F_{X_{11} X_{12}} + G^1_{\ \phi^2}  \right) \, .
\ee

Deriving this relation clarifies the meaning of the redundancy on the choice of off-shell relations that has been made explicit in the examples discussed above. We can understand the cases discussed above, for instance Eq.~\eqref{eq:thirdrel}, as particular cases of the partial differential equation~\eqref{eq:genoffc} in which different terms are required to vanish independently. Hence, a solution to this redundancy problem is working directly with solutions of Eq.~\eqref{eq:genoffc}.

Note that solutions to Eq.~\eqref{eq:genoffc} are guaranteed to exist, under the assumptions that the coupling functions are analytic and the initial data is non-characteristic, due to the Cauchy–Kovalevskaya theorem~\cite{sneddon2006elements,evans2010partial}. Moreover, this equation is of the hyperbolic type, as:
\begin{equation}
-4\mathcal{C}_1\mathcal{C}_2=16 X_{11}^6\left(G^1_{X_{11}}\right)^4\left(F+2G^1_{\phi^1}X_{11}\right)^2>0.    
\end{equation}
We have thus identified a well-posed formulation of the problem of finding field equations describing regular black holes with dynamical regularization scales. This formulation provides a clear starting point for more detailed explorations of the problem in which solutions of the equation above are studied.

As a final comment, we also note that there is a weaker formulation of the problem, in which Eq.~\eqref{eq:genoffc} is only implemented as an on-shell statement. While this approach can provide additional room for solutions, it also seems more difficult to analyze in practice. However, from an exploratory perspective, it is important to keep in mind that such an alternative route exists.

\section{Discussion \label{sec:discussion}}

In this article, we have proposed the idea of upgrading the regularization scale $\ell$ of regular black hole spacetimes to a dynamical field. Such an upgrade would, in principle, allow the regular core of said objects to evolve with time, coherently with what is suggested by the analysis of perturbations of static spacetimes~\cite{Barcelo:2020mjw,Barcelo:2022gii}. 

We have implemented this upgrade in the framework recently proposed in~\cite{Carballo-Rubio:2025ntd,Boyanov:2025pes} and further investigated in~\cite{BorissovaEffectiveGeometrodynamics2026,BorissovaRegularBlack2026a,BorissovaAll2D2026,BorissovaModifiedFriedmann2026,Arrechea:2026ngi,Borissova$g_ttg_rr1$2026,Carballo-Rubio:2026mvj,MazzaBlackBounces2026,BorissovaFormationExtremal2026,Thaalba:2026abz}, which allows to interpret regular black hole metrics as spherically symmetric solutions of a 2-dimensional Horndeski theory. In this context, the regularization scale typically appears as a coupling constant in the field equations, so that upgrading it to a field requires extending the formalism substantially. Since said framework is based on 2-dimensional Horndeski theory, its most natural extension consists in 2-dimensional biscalar Horndesky theory --- which is indeed the language we have adopted.

This article represents a first exploration into this subject. First of all, we have clarified how the field equations in~\cite{Carballo-Rubio:2025ntd} can be regarded as \qu{single-field} instances of biscalar Horndeski theories, and discussed what extending them actually means in this context. Notably, we have investigated a minimal strategy for implementing such an extension --- cf.~\cref{sec:singext}. Our exposition focused on static metric configurations, as these represent the direct analogs of the solutions of~\cite{Carballo-Rubio:2025ntd}; moreover, static solutions are supposed to act as equilibrium states that dynamical transients may connect. Then, we have made the simplifying assumption that the additional scalar field be constant, and analyzed this particular case in detail --- cf.~\cref{sec:const_scalar}. 
Specifically, we have discussed the identification of \qu{off-shell} constraints and reduced this problem to solving a linear partial differential equation for the new coupling function that is not present in the single-field case, as described in~\cref{sec:offshell}.

Along the way, we have also discussed specific subcases with the hope of finding analytical solutions. However, the additional assumptions required to define these subcases proved too stringent. We nevertheless decided to include these subcases both for completeness and as negative results that are useful for further explorations of the topic. These negative results provide additional motivation to consider the problem in its full generality and, in particular, its implementation in terms of the linear partial differential equation introduced in~\cref{sec:offshell}. While this approach is more complex, one of the main conclusions in our paper is that a more detailed treatment (either by analytical or numerical means) of this partial differential equation is required to reach a complete understanding of the question that motivated our study in the first place.

\acknowledgments
RCR acknowledges financial support provided by the Spanish Government through the Ram\'on y Cajal program (contract RYC2023-045894-I), the Grant No.~PID2023-149018NB-C43 funded~by MCIN/AEI/10.13039/501100011033, and the Severo Ochoa grant CEX2021-001131-S funded by MCIN/AEI/ 10.13039/501100011033. JM acknowledges support of ANR grant StronG (ANR-22-CE31-0015-01).

\appendix

\section{Derivation of the tensorial and scalar field equations}\label{app:tseqs}

We will start deriving the tensorial field equations by considering variations of the action for the Lagrangian in Eq.~\eqref{eq:bilag} with respect to $g^{ab}$. We will need the relations
\be
\var{X_{IJ}} = - \frac{1}{2} \phi^I_{(a} \phi^J_{b)}\var g^{ab}\, ,
\ee
and
\be
\var(\Box \phi^I) = \nabla_a \nabla_b \phi^I \var{g^{ab}} + \nabla_a \phi^I \nabla_b \var{g^{ab}} - \frac{1}{2} g_{ab} \nabla^c \phi^I \nabla_c \var{g^{ab}}\, .
\ee
Hence, we have
\be
\int \dd[2]{x} \delta\mathcal{L} &= \int \dd[2]{x} \sqrt{-g} \left\{
- \frac{1}{2} g_{ab}\left( F + G^I \Box \phi^I \right)\var{g^{ab}}
+ \left( F_{X_{IJ}} + G^K_{\ X_{IJ}} \Box \phi^K \right) \var{X_{IJ}}
\right.\nonumber\\
&\phantom{=} \left.
+ G^I \var{\Box \phi^I}
\right\} \\
&= \int \dd[2]{x} \sqrt{-g} \var{g^{ab}}\left\{
- \frac{1}{2} g_{ab}\left( F + G^I \Box \phi^I \right)
+ \left( F_{X_{IJ}} + G^K_{\ X_{IJ}} \Box \phi^K \right) \fdv{X_{IJ}}{g^{ab}}
\right. \nonumber\\
&\phantom{=} \left.
+ G^I \nabla_a \nabla_b \phi^I
- \nabla_{(a} \left[ G^I \nabla_{b)}\phi^I\right] + \frac{1}{2} g_{ab} \nabla_c \left( G^I \nabla^c \phi^I \right)
\right\} \\
&= \int \dd[2]{x} \sqrt{-g} \var{g^{ab}}\left\{
- \frac{1}{2} g_{ab} F - \frac{1}{2} \nabla_{(a} \phi^I \nabla_{b)} \phi^J \left( F_{X_{IJ}} + G^K_{\ X_{IJ}} \Box \phi^K \right)
\right. \nonumber \\
&\phantom{=} \left.
- \nabla_{(a} \phi^I \nabla_{b)} G^I + \frac{1}{2} g_{ab} \nabla^c \phi^I \nabla_c G^I
\right\} \\
&= \int \dd[2]{x} \sqrt{-g} \var{g^{ab}}\left\{
- \frac{1}{2} g_{ab} \left[ F + 2 X_{IJ} G^I_{\ \phi^J} + G^I_{\ X_{JK}} \nabla^c \phi^I \nabla^d \phi^{(J} \nabla_c\nabla_d \phi^{K)} \right]
\right. \nonumber \\
&\phantom{=} \left.
- \frac{1}{2} \nabla_{(a} \phi^I \nabla_{b)} \phi^J \left( F_{X_{IJ}} + G^K_{\ X_{IJ}} \Box \phi^K + 2 G^I_{\ \phi^J} \right)
\right. \nonumber \\
&\phantom{=} \left.
+ G^I_{\ X_{JK}} \nabla_{(a} \phi^I \nabla^c \phi^{(J} \nabla_{b)} \nabla_c \phi^{K)}
\right\} \, .
\ee
The field equations resulting from variations with respect to $g^{ab}$ can be then written as
\be\label{eq:app_tensorial_aux1}
\mathcal{E}_{ab} &:=g_{ab} \left( F + 2 G^I_{\ \phi^J} X_{IJ} \right) + \nabla_{(a} \phi^I \nabla_{b)} \phi^J \left( F_{X_{IJ}} + 2 G^I_{\ \phi^J} + G^K_{\ X_{IJ}} \Box \phi^K \right)
\nonumber \\
&\phantom{:=}
+G^I_{\ X_{JK}}\left[g_{ab} \nabla^c \phi^I \nabla^d \phi^{J} \nabla_c\nabla_d \phi^K  - 2  \nabla_{(a} \phi^I \nabla^c \phi^J \nabla_{b)} \nabla_c \phi^K\right]\nonumber\\
&\phantom{:}=0\,.
\ee
We can further simplify these equations by exploiting the fact that we are in 2 dimensions. As described in App.~\ref{app:2dide}, there are a number of relations that are valid for this specific dimensionality. Taking Eq.~\eqref{eq:1stid_sym} in particular, we notice that it can be rearranged to write
\begin{align}
&g_{ab}\nabla^c\phi^I \nabla^d\phi^J \nabla_c\nabla_d \phi^K - \nabla_{(a} \phi^I \nabla_c \phi^J \nabla_{b)} \nabla^c \phi^K \nonumber\\
=& - 2g_{ab} X^{IJ} \Box \phi^K + 2 X^{IJ} \nabla_a \nabla_b \phi^K - \nabla_{(a} \phi^I \nabla_{b)} \phi^J \Box \phi^K + \nabla_c \phi^I \nabla_{(a} \phi^J \nabla_{b)} \nabla^c \phi^K \, ,
\end{align}
and thus write
\be\label{eq:app_tensorial_aux2}
\mathcal{E}_{ab} &= g_{ab} \left( F + 2 G^I_{\ \phi^J} X_{IJ} - 2 G^I_{\ X_{JK}} X^{IJ} \Box \phi^K \right) 
\nonumber \\
&\phantom{=} + \nabla_{(a} \phi^I \nabla_{b)} \phi^J \left[ F_{X_{IJ}} + 2 G^I_{\ \phi^J} + \left( G^K_{\ X_{IJ}} - G^I_{\ X_{JK}} \right) \Box \phi^K \right]
\nonumber \\
&\phantom{:=}
+G^I_{\ X_{JK} } \left[2 X^{IJ} \nabla_a \nabla_b \phi^K + \nabla_c \phi^I \nabla_{(a} \phi^J \nabla_{b)} \nabla^c \phi^K - \nabla_{(a} \phi^I \nabla_c \phi^J \nabla_{b)} \nabla^c \phi^K \right]\,.
\ee
These are not the final expressions we will be using in the main body of the paper, as we will impose some conditions on the functions $G^I$ below when discussing the scalar field equations.

For variations of the action for the Lagrangian in Eq.~\eqref{eq:bilag} with respect to $\phi^I$, we have:
\be\label{eq:varscal}
\int \dd[2]{x} \var{\mathcal{L}} &= \int \dd[2]{x} \sqrt{-g} \left\{ 
\left(F_{\phi^I} + G^K_{\ \phi^I} \Box \phi^K \right) \var{\phi^I}
+ \left(F_{X_{IJ}} + G^K_{\ X_{IJ}} \Box \phi^K \right) \var{X_{IJ}}
+ G^K \var{\Box \phi^K}
\right\}
\nonumber\\
&= \int \dd[2]{x} \sqrt{-g} \left\{ 
\left(F_{\phi^I} + G^K_{\ \phi^I} \Box \phi^K \right) \var{\phi^I}
\right. \nonumber\\
&\phantom{=} \left.
+ \left(F_{X_{IJ}} + G^K_{\ X_{IJ}} \Box \phi^K \right) \left(-\frac{\nabla^a \phi^I \nabla_a \var{\phi^J} + \nabla^a \phi^J \nabla_a \var{\phi^I} }{2} \right)
+ (\Box G^K) \var{\phi^K}
\right\}
\nonumber\\
&= \int \dd[2]{x} \sqrt{-g} \left\{ 
F_{\phi^I} + G^K_{\ \phi^I} \Box \phi^K+ \nabla_a \left[\nabla^a \phi^J \left( F_{X_{IJ}} + G^K_{\ X_{IJ}} \Box \phi^K \right)\right]
+ \Box G^I
\right\}\var{\phi^I} 
\nonumber\\
&= \int \dd[2]{x} \sqrt{-g} \left\{ 
F_{\phi^I} + G^K_{\ \phi^I} \Box \phi^K 
+ \left( F_{X_{IJ}} + G^K_{\ X_{IJ}} \Box \phi^K \right) \Box \phi^J
\right. \nonumber\\
&\phantom{=} \left. 
\nabla^a \phi^J \left( \nabla_a F_{X_{IJ}} + \nabla_a G^K_{\ X_{IJ}} \Box \phi^K + G^K_{\ X_{IJ}} \nabla_a \Box \phi^K \right)
+ \Box G^I
\right\} \var{\phi^I}  \, .
\ee
Due to the length of the expressions involved, let us manipulate some of the terms in the equation above separately, starting with:
\be\label{eq:appaux1}
&\phantom{=}
\nabla^a \phi^J \left( \nabla_a F_{X_{IJ}} + \nabla_a G^K_{\ X_{IJ}} \Box \phi^K + G^K_{\ X_{IJ}} \nabla_a \Box \phi^K \right) = \nonumber\\
&=
\nabla^a \phi^J \left[ 
\left( F_{X_{IJ}\phi^K} + G^L_{\ X_{IJ} \phi^K} \Box \phi^L \right) \nabla_a \phi^K 
+ \left( F_{X_{IJ}X_{KL}} +  G^M_{\ X_{IJ} X_{KL}} \Box \phi^M\right) \nabla_a X_{KL} \right] 
\nonumber\\
&\phantom{=}
+ G^K_{\ X_{IJ}} \nabla^a \phi^J \nabla_a \Box \phi^K 
\nonumber\\
&= -2 X_{JK} \left( F_{X_{IJ}\phi^K} + G^L_{\ X_{IJ} \phi^K} \Box \phi^L \right)
+ \left( F_{X_{IJ}X_{KL}} +  G^M_{\ X_{IJ} X_{KL}} \Box \phi^M\right) \nabla^a \phi^J \nabla_a X_{KL}
\nonumber\\
&\phantom{=}
+ G^K_{\ X_{IJ}} \nabla^a \phi^J \nabla_a \Box \phi^K\,.
\ee
On the other hand,
\be
\Box G^I &= \nabla_a \left( G^I_{\ \phi^J} \nabla^a \phi^J + G^I_{\ X_{JK}} \nabla^a X_{JK}  \right)
\nonumber\\
&= G^I_{\ \phi^J} \Box \phi^J + G^I_{\ X_{JK}} \Box X_{JK}
+ \nabla^a \phi^J \left[ G^I_{\ \phi^J \phi^K} \nabla_a \phi^K + G^I_{\ \phi^J X_{KL}} \nabla_a X_{KL}  \right]
\nonumber\\
&\phantom{=}
+ \nabla^a X_{JK} \left[ G^I_{\ X_{JK} \phi^L} \nabla_a \phi^L + G^I_{\ X_{JK} X_{LM}} \nabla_aX_{LM}  \right]
\nonumber\\
&= G^I_{\ \phi^J} \Box \phi^J + G^I_{\ X_{JK}} \Box X_{JK} - 2 X_{JK} G^I_{\ \phi^J \phi^K}
\nonumber\\
&\phantom{=} + 2 G^I_{\ \phi^J X_{KL}} \nabla^a \phi^J \nabla_a X_{KL}
+ G^I_{\ X_{JK} X_{LM}} \nabla^a X_{JK} \nabla_a X_{LM} \, .
\ee
Using the following relation,
\be
\Box X_{JK} &= - \nabla^b \left[  \nabla^a \phi^{(J} \nabla_b \nabla_a \phi^{K)} \right]
\nonumber\\
&= - \left[  \nabla^a \phi^{(J} \Box \nabla_a \phi^{K)} + \nabla^a \nabla^b \phi^{(J} \nabla_a \nabla_b \phi^{K)} \right] \, ,
\ee
we can write
\be\label{eq:appaux2}
\Box G^I &= G^I_{\ \phi^J} \Box \phi^J - G^I_{\ X_{JK}} \left(  \nabla^a \phi^J \Box \nabla_a \phi^K + \nabla^a \nabla^b \phi^J \nabla_a \nabla_b \phi^K \right) - 2 X_{JK} G^I_{\ \phi^J \phi^K}
\nonumber\\
&\phantom{=} + 2 G^I_{\ \phi^J X_{KL}} \nabla^a \phi^J \nabla_a X_{KL}
+ G^I_{\ X_{JK} X_{LM}} \nabla^a X_{JK} \nabla_a X_{LM} \, .
\ee

We can not put together Eqs.~\eqref{eq:varscal},~\eqref{eq:appaux1} and~\eqref{eq:appaux2} to write the scalar field equation as
\be
\mathcal{E}^I &=\label{eq:eqqs3} F_{\phi^I} + G^K_{\ \phi^I} \Box \phi^K 
+ \left( F_{X_{IJ}} + G^K_{\ X_{IJ}} \Box \phi^K \right) \Box \phi^J
-2 X_{JK} \left( F_{X_{IJ}\phi^K} + G^L_{\ X_{IJ} \phi^K} \Box \phi^L \right)
\nonumber\\
&\phantom{=}
+ \left( F_{X_{IJ}X_{KL}} +  G^M_{\ X_{IJ} X_{KL}} \Box \phi^M\right) \nabla^a \phi^J \nabla_a X_{KL}
+ G^K_{\ X_{IJ}} \nabla^a \phi^J \nabla_a \Box \phi^K 
\nonumber\\
&\phantom{=}
+ G^I_{\ \phi^J} \Box \phi^J - G^I_{\ X_{JK}} \left(  \nabla^a \phi^J \Box \nabla_a \phi^K + \nabla^a \nabla^b \phi^J \nabla_a \nabla_b \phi^K \right) - 2 X_{JK} G^I_{\ \phi^J \phi^K}
\nonumber\\
&\phantom{=} + 2 G^I_{\ \phi^J X_{KL}} \nabla^a \phi^J \nabla_a X_{KL}
+ G^I_{\ X_{JK} X_{LM}} \nabla^a X_{JK} \nabla_a X_{LM} 
\nonumber\\
&= F_{\phi^I} + \left( G^J_{\ \phi^I} + G^I_{\ \phi^J} \right) \Box \phi^J  
+ \left( F_{X_{IJ}} + G^K_{\ X_{IJ}} \Box \phi^K \right) \Box \phi^J
\nonumber\\
&\phantom{=}
-2 X_{JK} \left( F_{X_{IJ}\phi^K} + G^I_{\ \phi^J \phi^K} + G^L_{\ X_{IJ} \phi^K} \Box \phi^L \right)
- G^I_{\ X_{JK}} \nabla^a \nabla^b \phi^J \nabla_a \nabla_b \phi^K
\nonumber\\
&\phantom{=}
+ \left( F_{X_{IJ}X_{KL}} + 2 G^I_{\ \phi^J X_{KL}} +  G^M_{\ X_{IJ} X_{KL}} \Box \phi^M\right) \nabla^a \phi^J \nabla_a X_{KL}
+ G^I_{\ X_{JK} X_{LM}} \nabla^\mu X_{JK} \nabla_\mu X_{LM}
\nonumber\\
&\phantom{=}
-G^I_{X_{JK}}R_{ab}\nabla^a\phi^J\nabla^b\phi^K+\left(G^K_{\ X_{IJ}}-G^I_{\ X_{JK}}\right) \nabla^a \phi^J \nabla_a \Box \phi^K
\, ,
\ee
where we have used the relation
\be
\Box \nabla_a \phi^K = \nabla_a \Box  \phi^K + R_{ab} \nabla^b \phi^K \, ,
\ee

The second term in the last line of Eq.~\eqref{eq:eqqs3} represents the only higher-derivative term in these equations. To remove the higher-derivative term, we need to impose
\be\label{eq:2ndordcond}
G^K_{\ X_{IJ}} = G^I_{\ X_{JK}} \, .
\ee
Hence, $G^K_{X_{IJ}}$ must be completely symmetric in its indices. This constraint was already discussed in~\cite{Nejati:2024tuo}. It follows that $G^K_{X_{IJ}X_{LM}}$ must be completely symmetric in its lower indices:
\begin{equation}\label{eq:2ndordcond2}
G^K_{X_{IJ}X_{LM}}=G^I_{X_{JK}X_{LM}}=G^I_{X_{LM}X_{JK}}=G^L_{X_{IM}X_{JK}}=G^L_{X_{JK}X_{IM}}=G^K_{X_{JL}X_{IM}},\end{equation}
and, as a consequence, in all its indices.

Assuming this condition is satisfied, the scalars' equations become
\be
\mathcal{E}^I &= 
 F_{\phi^I} + \left( G^J_{\ \phi^I} + G^I_{\ \phi^J} \right) \Box \phi^J 
+ \left( F_{X_{IJ}} + G^K_{\ X_{IJ}} \Box \phi^K \right) \Box \phi^J-G^I_{X_{JK}}R_{ab}\nabla^a\phi^J\nabla^b\phi^K
\nonumber\\
&\phantom{=}
-2 X_{JK} \left( F_{X_{IJ}\phi^K} + G^I_{\ \phi^J \phi^K} + G^L_{\ X_{IJ} \phi^K} \Box \phi^L \right)
- G^I_{\ X_{JK}} \nabla^a \nabla^b \phi^J \nabla_a \nabla_b \phi^K
\nonumber\\
&\phantom{=}
+ \left( F_{X_{IJ}X_{KL}} + 2 G^I_{\ \phi^J X_{KL}} +  G^M_{\ X_{IJ} X_{KL}} \Box \phi^M\right) \nabla^a \phi^J \nabla_a X_{KL}
+ G^I_{\ X_{JK} X_{LM}} \nabla^\mu X_{JK} \nabla_\mu X_{LM}
\nonumber\\
&= F_{\phi^I} 
+ 
\left(F_{X_{IJ}} + G^J_{\ \phi^I} + G^I_{\ \phi^J} \right) \Box \phi^J 
- G^K_{\ X_{IJ}} R_{ab} \nabla^a \phi^J \nabla^b \phi^K
-2 X_{JK} \left( F_{X_{IJ}\phi^K} + G^I_{\ \phi^J \phi^K} \right)
\nonumber\\
&\phantom{=}
G^K_{\ X_{IJ}} \Box \phi^J \Box \phi^K
- 2 G^L_{\ X_{IJ} \phi^K} X_{JK} \Box \phi^L
- G^I_{\ X_{JK}} \nabla^a \nabla^b \phi^J \nabla_a \nabla_b \phi^K
\nonumber\\
&\phantom{=}
+ \left( F_{X_{IJ}X_{KL}} + 2 G^I_{\ \phi^J X_{KL}} \right) \nabla^a \phi^J \nabla_b X_{KL}
\nonumber\\
&\phantom{=}
+ 
G^M_{\ X_{IJ} X_{KL}} \Box \phi^M \nabla^a \phi^J \nabla_a X_{KL}
+ G^I_{\ X_{JK} X_{LM}} \nabla^a X_{JK} \nabla_a X_{LM}
\nonumber\\
&= F_{\phi^I} 
+  
\left(F_{X_{IJ}} + G^J_{\ \phi^I} +  G^I_{\ \phi^J}\right) \Box \phi^J 
-2 X_{JK} \left( F_{X_{IJ}\phi^k} + G^I_{\ \phi^J \phi^K} - \frac{1}{2} R G^K_{\ X_{IJ}} \right)
\nonumber\\
&\phantom{=}
+ 
G^I_{\ X_{JK}} \left( \Box \phi^J \Box \phi^K
- \nabla^a \nabla^b \phi^J \nabla_a \nabla_b \phi^K \right)
- 2 G^I_{\ \phi^J X_{KL}} \left(  X_{JK} \Box \phi^L - \nabla^a \phi^J \nabla_a X_{KL} \right)
\nonumber\\
&\phantom{=}
+ F_{X_{IJ}X_{KL}}  \nabla^a \phi^J \nabla_a X_{KL}
+ G^I_{\ X_{JK} X_{LM}} \left(
\Box \phi^M \nabla^a \phi^J \nabla_a X_{KL}
+ \nabla^a X_{JK} \nabla_a X_{LM}
\right)=0
\, .
\ee
In the last line, we have used explicitly the symmetry properties in Eqs.~\eqref{eq:2ndordcond} and~\eqref{eq:2ndordcond2} to regroup some of the terms. Rearranging the terms and using Eq.~\eqref{eq:2ndid}, we arrive at the equivalent expression
\be\label{eq:app_scalar}
\mathcal{E}^I &= F_{\phi^I} 
+  
\left(F_{X_{IJ}} + G^J_{\ \phi^I} +  G^I_{\ \phi^J}- 2 G^I_{\ \phi^L X_{KJ}}  X_{LK}\right) \Box \phi^J \nonumber\\
&\phantom{=}-2 X_{JK} \left( F_{X_{IJ}\phi^K} + G^I_{\ \phi^J \phi^K} - \frac{1}{2} R G^K_{\ X_{IJ}} \right)+ \left(F_{X_{IJ}X_{KL}}  +2 G^I_{\ \phi^J X_{KL}}\right)\nabla^a \phi^J \nabla_a X_{KL}
\nonumber\\
&\phantom{=}
+ 
G^I_{\ X_{JK}} \left( \Box \phi^J \Box \phi^K
- \nabla^a \nabla^b \phi^J \nabla_a \nabla_b \phi^K \right)
\nonumber\\
&\phantom{=}
+ G^I_{\ X_{JK} X_{LM}} \left(
\Box \phi^M \nabla^a \phi^J \nabla_a X_{KL}
+ \nabla^a X_{JK} \nabla_a X_{LM}
\right)=\nonumber\\
&= F_{\phi^I} 
+  
\left(F_{X_{IJ}} + G^J_{\ \phi^I} +  G^I_{\ \phi^J}- 2 G^I_{\ \phi^L X_{KJ}}  X_{LK}\right) \Box \phi^J \nonumber\\
&\phantom{=}-2 X_{JK} \left( F_{X_{IJ}\phi^k} + G^I_{\ \phi^J \phi^K} - \frac{1}{2} R G^K_{\ X_{IJ}} \right)+ \left(F_{X_{IJ}X_{KL}}  +2 G^I_{\ \phi^J X_{KL}}\right)\nabla^a \phi^J \nabla_a X_{KL}
\nonumber\\
&\phantom{=}
+ 
\left(G^I_{\ X_{JK}} +X_{LM}G^I_{\ X_{JK} X_{LM}}\right)\left( \Box \phi^J \Box \phi^K
- \nabla^a \nabla^b \phi^J \nabla_a \nabla_b \phi^K \right)=0
\, .
\ee

On the other hand, under the conditions~\eqref{eq:2ndordcond}, the tensorial field equations become
\be\label{eq:app_tensorial}
\mathcal{E}_{ab} &=  \left( F + 2 G^I_{\ \phi^J} X_{IJ} \right)g_{ab}+2 X^{IJ}G^I_{\ X_{JK} }\left( \nabla_a\nabla_b\phi^K-g_{ab}\square\phi^K \right)
\nonumber \\
&\phantom{=} +  \left(F_{X_{IJ}}+ 2 G^I_{\ \phi^J} \right)\nabla_{(a} \phi^I \nabla_{b)} \phi^J=0\,.
\ee

\section{2-dimensional identities}\label{app:2dide}

In this appendix, we derive some identities that are strictly valid in 2 dimensions, although similar relations exist in higher dimensions \cite{LovelockDimensionallyDependent1970}. 
All of them descend from properties of generalized Kronecker deltas (see e.g.~\cite{Frankel:1997ec} for a definition), and specifically by noticing that all generalized Kronecker deltas with more than 2 indices must vanish, in particular
\be
\delta^{abc}_{def} = 0\,.
\ee

This allows us to obtain an identity of tensorial rank 2, that is cubic in the fields $\phi^I$, and quartic in derivatives:
\be\label{eq:1stid}
0 &= \delta^{a c d}_{b e f } \nabla_c \phi^I \nabla_e \phi^J \nabla_d \nabla^f \phi^K \nonumber\\
&= - \delta^a_b \left( 2X^{IJ} \Box \phi^K + \nabla^c\phi^I \nabla^d\phi^J \nabla_c\nabla_d \phi^K \right) + 2 X^{IJ} \nabla^a \nabla_b \phi^K 
\nonumber\\
&\phantom{=}
- \nabla_b \phi^I \nabla^a \phi^J \Box \phi^K + \nabla_c \phi^I \nabla^a \phi^J \nabla_b \nabla^c \phi^K + \nabla_b \phi^I \nabla_c \phi^J \nabla^a \nabla^c \phi^K \, ;
\ee
Lowering an index, we notice that the first line is symmetric in $a$ and $b$, and therefore the second line must be also. We thus get
\be\label{eq:1stid_sym}
0 &= - g_{ab} \left( 2X^{IJ} \Box \phi^K + \nabla^c\phi^I \nabla^d\phi^J \nabla_c\nabla_d \phi^K \right) + 2 X^{IJ} \nabla_a \nabla_b \phi^K 
\nonumber\\
&\phantom{=}
- \nabla_{(a} \phi^I \nabla_{b)} \phi^J \Box \phi^K + \nabla_c \phi^I \nabla_{(a} \phi^J \nabla_{b)} \nabla^c \phi^K + \nabla_{(b} \phi^I \nabla^c \phi^J \nabla_{a)} \nabla_c \phi^K \, , \\
0&= - \nabla_{[a} \phi^I \nabla_{b]} \phi^J \Box \phi^K + \nabla_c \phi^I \nabla_{[a} \phi^J \nabla_{b]} \nabla^c \phi^K + \nabla_{[b} \phi^I \nabla^c \phi^J \nabla_{a]} \nabla_c \phi^K \, ;
\ee
the first of these identities has been used in \cref{app:tseqs} to pass from \cref{eq:app_tensorial_aux1} to \cref{eq:app_tensorial_aux2}.

From \cref{eq:1stid}, one can obtain further identities by contracting its tensorial indices with, say, $\nabla^b \phi^L$ (tensorial rank 1, 4 fields, 5 derivatives), and $\nabla^a \phi^M \nabla^b \phi^L$ (tensorial rank 0, 5 fields, 6 derivatives). These identies were not used in the body of the text, and for this reason we omit them.

Yet another, independent identity is of tensorial rank 1, cubic in the fields, and containing 5 derivatives:
\be
0 &= \delta^{abc}_{def} \nabla_b \nabla^d \phi^I \nabla_c \nabla^e \phi^J \nabla^f \phi^K \nonumber\\
&= \nabla^a \phi^K \left[ \Box \phi^I \Box \phi^J - \nabla^b \nabla^c \phi^I \nabla_b \nabla_c \phi^J \right]
+ 2 \nabla^a \nabla^b \phi^{(I} \nabla_b \nabla_c \phi^{J)} \nabla^c \phi^K
\nonumber\\
&\phantom{=}
- 2 \nabla^a \nabla^b \phi^{(I} \Box \phi^{J)} \nabla_b \phi^K 
\, .
\ee
Contracting this with $\nabla_a \phi^L$, we derive the following (tensorial rank 0, 4 fields, 6 derivatives):
\be
0 &= - 2 X_{KL} \left[ \Box \phi^I \Box \phi^J - \nabla^a \nabla^b \phi^I \nabla_a \nabla_b \phi^J \right]
+ 2 \nabla_a \phi^L \nabla^a \nabla^b \phi^{(I} \nabla_b \nabla_c \phi^{J)} \nabla^c \phi^K
\nonumber\\
&\phantom{=}
- 2 \nabla_a \phi^L \nabla^a \nabla^b \phi^{(I} \Box \phi^{J)} \nabla_b \phi^K 
\, ;
\ee
when the internal indices $I$, $J$, $K$, and $L$ are contracted with an object $S_{(IJKL)}$ that is completely symmetric, the identity above takes up a more compact form:
\be\label{eq:2ndid}
0 
&= S_{(IJKL)} \left\{ 
X_{KL} \left[ \Box \phi^I \Box \phi^J - \nabla^a \nabla^b \phi^I \nabla_a \nabla_b \phi^J \right]
- \nabla^a X_{IK} \left[ \nabla_a X_{JL} + \nabla_a \phi^J \Box \phi^L \right]
\right\} \, .
\ee
This identity has been used in \cref{app:tseqs}, to pass from the first to the second identity in \cref{eq:app_scalar} (recall that $G^I_{\ X_{JK}X_{LM}}$ is symmetric in all of its indices).


\bibliographystyle{JHEP}
\bibliography{biblio.bib}

\end{document}